# Obtaining the Specific Heat of Hadronic Matter from CERN/ RHIC Experiments

Aram Mekjian

Rutgers University, Department of Physics, Piscataway, N.J. 08854

## Abstract

The specific heat of hot hadronic matter is related to particle production yields from experiments done at CERN/RHIC. The mass fluctuation of excited hadrons plays an important role. Connections of the specific heat, mean hadronic mass excited and its fluctuation with properties of baryon and electric charge chemical potentials (value, slope and curvature) are also developed. A possible divergence of the specific heat as $1/(T_0 - T)^2$ is discussed. Some connections with net charge fluctuations are noted.

## 1. Introduction

The statistical model of very high energy collisions can account for particle production yields from very high energy collisions [1-5]. The same model also contains information regarding the thermodynamic proper ties of this system of particles. One important thermodynamic property is the specific heat. In this paper, expressions for the specific heat will first be developed. The importance of a study of the specific heat stems from the fact that sudden changes in the specific heat have been used as signals for phase transitions. A classic example of this statement is the lambda transition in liquid helium. The name lambda transition reflects the lambda shape of the specific heat with a very sharp rise followed by a sudden decrease. In the liquid-gas phase transition of nuclear matter at moderate excitation energy or temperature a very similar sharp peak in the specific heat was found in a theoretical model developed in ref [6,7]. This is associated with the increase in the surface energy of the system as the original nucleus breaks into small and smaller clusters with increasing temperature. For the situation discussed here, a rapid rise in the specific heat is associated with large fluctuations in the mass spectrum of the excited particles. Event-by event studies [8] have also been stressed along with temperature fluctuations [9]. Large values of the specific heat are associated with large energy fluctuations. The compressibility is associated with density fluctuations [10]. Fluctuations associated with net electric charge and baryon charge have also been of recent interest [11,12] as well as $p_t$ fluctuations [13]. An overview of fluctuations and correlations can be found in [14,15]. Connections with some of the quantities that appear in this paper with baryon and electric charge fluctuations will be mentioned.

The organization of this paper is as follows. First, results of the statistical model are given for particle production yields and for thermodynamic quantities in situations where constraints associated with conservation laws are important such as in heavy ion collisions. The specific heat is then connected to properties of the particle production yields and conserved charges such as baryon number $B$ and electric charge $Z$. Limiting cases of the specific heat are discussed which show a connection of the specific heat to the mass fluctuation in the spectrum of excited particles. Connections of the specific heat with the behavior of the chemical potential (its value, slope and curvature) are also developed. The distribution of particles obtained from the detailed analysis of fitting the statistical model to hadronic multiplicities in Pb-Pb collisions at 30A, 40A and 80A GeV data [3] is then used to study the behavior of the specific heat. A parameterization of the behavior of the chemical potential with $T$ from this analysis may indicate a sharp increase in the specific heat.

## 2. Statistical Model.
### 2.1 Statistical and thermodynamic properties.

The statistical model of heavy ion collisions assumes that hadron multiplicities are the result of an established thermal and chemical equilibrium [16] in some interaction volume $V$ at some temperature $T$. The interaction volume is the freeze out volume which is the largest volume over which equilibrium is maintained in the evolution of the fireball. The statistical equilibrium is developed from the underlying collisions between particles from the strong force. Here initial reaction rates are assumed fast enough compared to an expansion rate so that a quasi-equilibrium can be achieved. As the system expands, reaction rates quickly drop because of rapidly decreasing densities and equilibrium is broken at some point in the evolution. The simplest assumption is that all particles freeze out at the same volume $V$ and temperature $T$. The particle multiplicity distributions is model are then

$$<N_j> = g_j (Vm_j^2 T/2\pi^2) \sum_k K_2(km_j/T) \, Exp[b_j \, \mu_B \, k/T + q_j \, \mu_Q \, k/T]/k \qquad (1)$$

The $b_j$ and $q_j$ are the baryon number and charge of particle $j$ which has spin degeneracy $g_j$ and mass $m_j$. The $\mu_B$ and $\mu_Q$ are the baryon and charge chemical potentials. The strangeness chemical potential $\mu_S$ will be set equal to 0 and the strangeness suppression factor $\gamma_S$ will be set equal to 1. The main focus will be on the baryon and electric charge conservation in systems with large $B$ and $Z$. The energy of particle $j$ is given by

$$<E_j> = g_j (Vm_j^3 T/2\pi^2) \sum_k ((3/4)K_3 + (1/4)K_1) \, Exp(k(b_j \, \mu_B + q_j \, \mu_Q)/T)/k \qquad (2)$$

The arguments of the Bessel $K$ functions in eq(2) are the same as in eq(1). The energy equation has the particles rest mass within it. The sum over $k$ in the above equations gives the degeneracy corrections, with the $k=1$ term the non-degenerate or Maxwell Boltzmann limit. For non-relativistic particles in the non-degenerate limit, the

$$<N_j> = g_j (V/\lambda_j^3) \, Exp((-m_j + b_j \, \mu_B + q_j \, \mu_Q)/T) \qquad (3)$$

The thermal wavelength $\lambda$ of particle $j$ is given by $\lambda_j = h/(2\pi m_j T)^{1/2}$. The energy in this limit is simply $<E_j> = <N_j>(m_j + (3/2)T)$. For zero mass particles: $<N_j> = g_j(VT^3/\pi^2) \, Exp(b_j \, \mu_B + q_j \, \mu_Q)/T)$, $<E_j> = 3T <N_j>$.

## 2.2 Features of the chemical potentials $\mu_B$ and $\mu_Q$.

The two chemical potentials are determined by the constraint conditions on total baryon number $B$ and total charge $Q$ or $Z$. Namely: $B = \Sigma_j b_j <N_j>$ and $Z = \Sigma_j q_j <N_j>$. Moreover, the derivative of these chemical potentials with respect to T can be obtained from the conditions $\partial B/\partial T = 0$ and $\partial Z/\partial T = 0$. Also, use will later be made of $\partial^2 B/\partial T^2 = 0$ and $\partial^2 Z/\partial T^2 = 0$. As an example, consider a system with the multiplicity of all particles given by the non-degenerate non-relativistic limit of eq(3). Then,

$$(3B/2) + M_B/T + C_{bb} f_B + C_{bq} f_Q = 0 \tag{6}$$
$$(3Z/2) + M_Q/T + C_{bq} f_B + C_{qq} f_Q = 0$$

Here, the various quantities that appear in eq(6) are defined by the following equations:

$$M_B = \sum_j m_j b_j N_j, \qquad M_Q = \sum_j m_j q_j N_j$$
$$C_{bb} = \sum_j b_j^2 N_j, \quad C_{qq} = \sum_j q_j^2 N_j, \quad C_{bq} = \sum_j b_j q_j N_j \tag{7}$$
$$f_B = \partial \mu_B/\partial T - \mu_B/T, \quad f_Q = \partial \mu_Q/\partial T - \mu_Q/T$$

In the above equations $N_j = <N_j>$. The coupled equations for $f_B$ and $f_Q$ give

$$f_B = (-(3B/2 + M_B/T)C_{qq} + (3Z/2 + M_Q/T)C_{bq})/C$$
$$f_Q = (-(3Z/2 + M_Q/T)C_{bb} + (3B/2 + M_B/T)C_{bq})/C \tag{8}$$

where $C = C_{bb} C_{qq} - C_{bq}^2$. If we neglect antibaryons production and take $b_j = 1$ for all baryons (no composites) then $C_{bb} = B$. Mesons don't contribute to either $C_{bb}$ or $C_{bq}$. Antiparticles enhance $C_{bb}$, $C_{qq}$ and contribute to $C_{bq}$ with the same sign as the associated particle. It is important to note that $C_{bb}$ and $C_{qq}$ depend on $b_j^2$ and $q_j^2$. They are measures of baryon number and charge fluctuations and depend on the fundamental baryon number and electric charge. In a Q-g phase these coefficients will be different since the unit of charge is 1/3 rather than 1 as used in ref [11,12]. Here, the focus will be properties of the hadron phase and the rise of the specific heat as it approaches a possible transition temperature $T_0$. In a future paper a discussion of the specific heat of the Q-g phase will be given. It can be calculated in the simple approximation of ideal gases of gluons and quarks using $T \partial S/\partial T$, where $S$ is the entropy. For the case of massless pions and all other mesons and baryons taken in the non-relativistic limit, the results of eq(8) become

$$f_B = (-(3B/2 + M_B/T)C_{qq} + (3Z/2 + 3Z_\pi/2 + M_Q/T)C_{bq})/C$$
$$f_Q = (-(3Z/2 + 3Z_\pi/2 + M_Q/T)C_{bb} + (3B/2 + M_B/T)C_{bq})/C \tag{9}$$

The $Z_\pi = N_{\pi^+} - N_{\pi^-}$, while the $Z$ is the total conserved charge and also contains the contribution of the charged pions. However, massless pions do not affect $M_Q$. For non-relativistic pions, the contribution of pions appears in $M_Q$ and the extra $Z_\pi$ term in eq(9) is no longer present.

## 2.3 Expressions for the specific heat of hadronic matter.

The specific heat of hadronic matter in the non-degenerate and the non-relativistic limit for all particles is given by

$$C_V = (\frac{3}{2})\sum_j N_j + \sum_j (\frac{3}{2} + \frac{m_j}{T})^2 N_j +$$
$$\{-(\frac{3B}{2} + \frac{M_B}{T})^2 C_{qq} - (\frac{3Z}{2} + \frac{M_Q}{T})^2 C_{bb} + 2(\frac{3B}{2} + \frac{M_B}{T})(\frac{3Z}{2} + \frac{M_Q}{T})C_{bq}\}/C \quad (10)$$
$$= (\frac{3}{2})\sum_j N_j + \sum_j (\frac{3}{2} + \frac{m_j}{T})^2 N_j - C_{bb}(f_B)^2 - C_{qq}(f_Q)^2$$

The first term on the right hand side of the last equation is just the ideal gas specific heat of each non-relativistic particle, with both mesons and baryons contributing. The second term involves the mass spectrum of all particles produced. The curly bracket or third term in the first equality has three contributions and involves the three coefficients $C_{bb}$, $C_{qq}$ and $C_{bq}$. The second and the third term arise from the possibility that the particle distributions change with $T$ or $\partial N_j/\partial T \neq 0$. The second term will be cancelled by the third term for a system which has $\partial N_j/\partial T = 0$, for all $j$. Specifically, consider a system of $N_p$ protons and $N_n$ neutrons, so that $N_p = Z$ and $N_p + N_n = B$. Then $C_{bb} = B$, $C_{qq} = Z$ and $C_{bq} = Z$. When these results are substituted into eq(12), along with the results $M_B = m_N B$ and $M_Q = m_N Z$, the second and the third term exactly cancel. The specific heat then reduces to the ideal gas law $C_V = (3/2)B$. The last equality replaces the curly bracket term with the functions $-C_{bb}(f_B)^2 - C_{qq}(f_Q)^2$ and connects $C_V$ to information about the behavior of the baryon and electric charge chemical potentials.

When pions are taken in the zero mass and non-degenerate limit, the $C_V$ is somewhat modified and now reads

$$C_V = (\frac{3}{2})\sum_j N_j + \sum_j (\frac{3}{2} + \frac{m_j}{T})^2 N_j + 12(N_{\pi^+} + N_{\pi^-} + N_{\pi^0}) +$$
$$\{-(\frac{3B}{2} + \frac{M_B}{T})^2 C_{qq} - (\frac{3Z}{2} + \frac{3Z_\pi}{2} + \frac{M_Q}{T})^2 C_{bb} + 2(\frac{3B}{2} + \frac{M_B}{T})(\frac{3Z}{2} + \frac{3Z_\pi}{2} + \frac{M_Q}{T})C_{bq}\}/C \quad (11)$$

The first two sums over $j$ on the right hand side of this equation exclude the pion contribution in their evaluation. The pion contribution is now contained in the following terms in that equation. The third term is the direct contribution of the pion as if it were independent of the charge conservation law and the curly bracket term arises from the chemical potentials and associated constraints. When these constraints are neglect, the curly bracket term is zero. The independent pion contribution can also be calculated using the results of eq(1) and eq(2). The exact expression for the specific heat per particle of an unconstrained meson or boson including statistical corrections reads:

$$C_{V,m}/N_m = (m/T)^2 \{ \sum_k K_2(km/T) + 3(T/m)K_3(km/T)/k \} / \sum_k K_2(km/T)/k \qquad (12)$$

The $m \to 0$ limit of this equation is $C_{V,m}/N_m = 12\varsigma(4)/\varsigma(3)$. If statistical corrections are neglected this limit would be 12, with the zeta functions giving the corrections from the sums over $k$ in eq(12). The non-degenerate and large $m/T$ limit of eq(12) is

$$C_{V,m}/N_m = (3/2 + m/T)^2 + 3/2 \qquad (13)$$

This is the characteristic dependence of the first 2 terms in eq(10).

### 2.4 Simplified model 1; only conserved baryon charge

To gain some further insight into properties of $C_V$, a simplified situation of one conserved charge will be considered. Namely, in this subsection only baryon number conservation will be imposed on the system. Then all charge and neutral states of the same baryon will have equal yields. Mesons and baryons will also completely decouple and the specific heat $C_V$ will be a sum of independent contributions from mesons given by eq(12) and constrained baryons. For this model, the baryon constraint of eq(6) is $3B/2 + M_B/T + C_{bb} f_B = 0$ and the contribution of the baryons to the specific heat, $C_{V,B}$:

$$C_{V,B} = \frac{3}{2} \sum_j N_{j,B} + \sum_j (\frac{3}{2} + \frac{m_j}{T})^2 N_{j,B} - (\frac{1}{C_{bb}})(\frac{3B}{2} + \frac{M_B}{T})^2 \qquad (14)$$

The sums over $j$ are over both baryons and anti-baryons. Anti-baryons are usually a small fraction of the total baryon number. If we allow only baryons with $b_j = 1$, then $C_{bb} = B$, and eq(19) can be rewritten in a simpler form

$$C_{V,B} = \frac{3}{2} B + B \frac{1}{T^2} (<m_B^2> - <m_B>^2) \qquad (15)$$

where the mean mass and its fluctuation are determined by

$$<m_B> = \sum_j m_j N_{j,B}/B, \qquad <m_B^2> = \sum_j m_j^2 N_{j,B}/B \qquad (16)$$

Thus the enhancement to the ideal gas specific heat $C_{V,B} = (3/2)\,B$ involves the mean square fluctuation in the resonance mass excitation. The $f_B$ can be used to obtain the mean baryonic mass that is excited by rearranging the baryon constraint condition to read

$$<m_B> = -3T/2 + (C_{bb}/B)(\mu_B - \partial\mu_B/\partial T) \tag{17}$$

Again, if anti-baryon production is neglected, the coefficient $C_{bb}/B = 1$. The condition $\partial^2 B/\partial T^2 = 0$ can be used to obtain an expression for the mean square fluctuation in the masses that are excited. This condition and the case for all $b_j = 1$ gives

$$<m_B^2> - <m_B>^2 = -3T^2/2 - T^3 \partial^2 \mu_B/\partial T^2 \tag{18}$$

Using this last result, the $C_{V,B}$ is simply

$$C_{V,B} = -T\partial^2 \mu_B/\partial T^2 = 3/2 + (\delta m_B)^2/T^2 \tag{19}$$

### 2.5 Role of anti-baryons

The presence of anti-baryons will modify some of the results given in sect[2.4]. For collision energies $\leq 100A\,GeV$ anti-baryons make up a few percent of $B$. From ref(3), the anti-proton, proton ratio is ~2% for the $80A\,GeV$ $Pb + Pb$ collision. This ratio rises to ~5% for the 158A $GeV$ collision. For a $\mu_B \sim 300\,MeV$ and $T \sim 150$, $\exp(-2\mu_B/T) = \exp(-4.) \sim 2\%$, which determines the anti-particle/ particle ratio in the absence of an electric chemical potential. The anti-particle, particle ratio will increase at much higher energy because $\mu_B$ decreases and $T$ increases. The presence of both anti-baryons with fraction $y = N_{\bar{B}}/(N_B + N_{\bar{B}})$ and baryons with fraction $x = N_B/(N_B + N_{\bar{B}})$ leads to a modified form for the baryonic (plus anti-baryonic) $C_{V,B}$:

$$\begin{aligned} C_{V,B}/(N_B+N_{\bar{B}}) &= \\ &3/2 + (x(\delta m_B)^2 + y(\delta m_{\bar{B}})^2)/T^2 + xy(3 + (<m_B> + <m_{\bar{B}}>)/T)^2 \\ &= 3/2 + (\delta m_B)^2/T^2 + 4xy(3/2 + <m_B>/T)^2 \end{aligned} \tag{20}$$

The last equality arises from $<m_B> = <m_{\bar{B}}>$ and $\delta m_B = \delta m_{\bar{B}}$. The specific heat is also

$$\begin{aligned} C_{V,B}/(N_B+N_{\bar{B}}) &= -((N_B+N_{\bar{B}})/B)(T\partial^2\mu_B/\partial T^2) + 4(N_B N_{\bar{B}}/B^2)f_B^2 \\ &= -\coth(\mu_B/T)T\partial^2\mu_B/\partial T^2 + 2\operatorname{csch}^2(\mu_B/T)(\partial\mu_B/\partial T - \mu_B/T)^2 \end{aligned} \tag{21}$$

When $N_{\bar{B}} \to 0$, or $\mu_B/T$ very large, the rhs of eq(20,21) reduces to eq(19). The specific heat now involves both the curvature and slope of the chemical potential and the value of the chemical potential itself. In the limit $x = y = 1/2$ the $C_{V,B}$ of eq(20) becomes the unconstrained limit: $C_{V,B}/(N_B+N_{\bar{B}}) = 3/2 + (9/4 + 3<m_B>/T + <m_B^2>/T^2$. The ideal gas limit $C_{V,B}/B = 3/2$ is realized in the limit $x=1$, $y=0$ and $\delta m_B = 0$. The $<m_B>$ and $<m_B^2> - <m_B>^2 = (\delta m_B)^2$ are given by

$$<m_B>/T = -3/2 - \coth(\mu_B/T)(\partial \mu_B/\partial T - \mu_B/T) \tag{22}$$
$$(\delta m_B)^2/T^2 = -3/2 - \coth(\mu_B/T)\cdot(T\partial^2 \mu_B/\partial T^2) + \text{csch}^2(\mu_B/T)(\partial \mu_B/\partial T - \mu_B/T)^2$$

Thus, the values of $\mu_B$, its derivative and curvature also contain the information necessary to evaluate various quantities of interest regarding the mass excitation.

## 2.6 Role of electric charge conservation

A hybrid case consists of no anti-baryon production (all $b_j = 1$), and having both baryon and electric charge conservation. This model reflects the fact that anti-baryon production is suppressed compared to $\pm$ charged particle production. The production of $\pm$ charged pion pairs is easier than a baryon-antibaryon pair. This hybrid model is also useful as a way of seeing how the two constraints act together in the expression for the specific heat and mass spectrum of produced hadrons. To keep final results as simple as possible within this hybrid case, the assumptions $<m_{B,+Q}> = <m_{B,-Q}> = <m_{B,0}>$ and $<m_{M,+Q}> = <m_{M,-Q}>$ will be made. Here, the subscripts ($B,+Q$) refer to baryons with charge $+Q$, ($M,+Q$) to mesons with $+Q$, etc. The uncharged mesons, ($M,0$), are decoupled from the baryon and electric charge conservation conditions and consequently add to $C_V$ independently. $C_V$ in this case reads

$$C_V = (\frac{3}{2})(B+N_M) + (\frac{3}{2} + \frac{<m_{B,+Q}>}{T})^2 B + (\frac{3}{2} + \frac{<m_{M,+Q}>}{T})^2 N_M +$$
$$\frac{B\cdot \delta m_B^2 + N_M \cdot \delta m_M^2}{T^2} - B f_B^2 - (N_+ + N_-)f_Q^2 \tag{23}$$

The $B = N_{B,+Q} + N_{B,-Q} + N_{B,0}$, $N_M = N_{M,0} + N_{M,+Q} + N_{M,-Q}$, $N_\pm$ is the total $\pm$ charge in mesons and baryons or $N_\pm = N_{M,\pm Q} + N_{B,\pm Q}$ and $N_+ + N_- = N_{CB} + N_{CM}$ with $N_{CB}$ and $N_{CM}$ the total number of charged baryons and charged mesons respectively. The mean baryon mass and mean meson mass are given by

$$<m_{B,+Q}>/T = -3/2 - f_B - (Z_B/B)f_Q \tag{24}$$
$$<m_{M,+Q}>/T = -3/2 - (1/Z_M)(N_+ + N_- - (Z-Z_M)^2/B))f_Q$$

The $Z = Z_M + Z_B$, where $Z_M = N_{M,+Q} - N_{M,-Q}$ and $Z_B = N_{B,+Q} - N_{B,-Q}$. Also

$$(\delta m_{B,+Q})^2/T^2$$

$$= -3/2 - T\partial^2 \mu_B/\partial T^2 - (Z_B/B)T\partial^2 \mu_Q/\partial T^2 - (N_{B,+Q} + N_{B,-Q} - (Z_B^2/B)) f_Q^2 \quad (25)$$

for the mean square fluctuation for baryons and the mean square fluctuation for mesons is

$$Z_M (\delta m_M)^2 / T^2 = -3Z_M/2 - (N_+ + N_- - Z_B^2/B)T\partial^2 \mu_Q/\partial T^2 - a_Q f_Q^2 \quad (26)$$

with $a_Q$ given by

$$a_Q = Z - 3Z_B N_{CB}/B + Z_B^3/B^2 + (1/Z_M)((N_{CB} - Z_B^2/B)^2 - N_{CM}^2) \quad (27)$$

The specific heat can then be obtained from the above equations when they are substituted into eq(23). The result relates $C_V$ to properties of $\mu_B$ and $\mu_Q$. The curvature of $\mu_B$ and $\mu_Q$ appear in expressions for the mass fluctuation, which also involves the slope and value of these functions. The mean excited meson mass only involves properties of $\mu_Q$ such as its slope and value, while the mean excited baryon mass involves similar properties of both $\mu_B$ and $\mu_Q$.

2.6 Some features of the statistical model analysis of CERN/RHIC data.

The result of the previous sections can be used to evaluate some features of recent CERN/RHIC data. In this ref[3], the $T$ dependence of $\mu_B$ is given in an equation that reads $T = .167 - .153\mu_B^2$. The $T$ and $\mu_B$ are in $GeV$. Here, a more general form for the $T$ dependence of $\mu_B$ is used, namely $T = T_0 - a(\mu_B)^{1/\beta}$. Substituting this result into eq(21) and eq(23) gives for $T \to T_0$

$$<m_B>/T \to \beta \cdot T_0/(T_0 - T), \quad (\delta m_B)^2/T^2 \to \beta \cdot T_0^2/(T_0 - T)^2$$
$$C_{V,B}/(N_B + N_{\bar{B}}) \to \beta(\beta+1) \cdot T_0^2/(T_0 - T)^2 \quad (28)$$

Thus, $\beta$ does not appear as an exponent in the divergence of the specific heat which is quadratic or $1/(T_0 - T)^2$ for all $\beta$. The $<m_B> \to \infty$ as $T \to T_0$. Using values for $\mu_B$ and $T$ given in ref[3] for $Pb + Pb$ collisions at 30, 40, 80 and 158A $GeV$ and for $Au + Au$ 11.6A $GeV$ collisions, and the above parameterization of the behavior of $\mu_B$ with $T$ with $\beta = 1/2$, the values of $<m_B>$, $(<m_B^2> - <m_B>^2)^{1/2} \equiv \delta m_B$ in $MeV$ and $C_{V,B}/B$ are shown in Table1. The last three columns are obtained from results in sect[2.5] that include antibaryons. The previous three columns are without anti-baryons obtained from expressions in sect[2.4]. The curvature and slope functions are obtained from this parameterization. The anti-baryon case also used this parameterization to evaluate the chemical potential, while the case with just baryons used the chemical potential of ref(3). The error bars in $T$ and $\mu_B$ are not given and generate large error bars in the results for the mean mass, mass fluctuation and specific heat, especially at the higher temperatures. These errors are typically $\pm 20\%$. Two sets of numbers for each energy appear in the table since ref[3] has two main analysis of the data, called A and B. The results presented in the table show that the specific heat per particle for baryons are very different from the ideal gas contribution of 1.5.

Table1. Values of the specific heat per particle, mean baryonic mass $<m_B>$ and mass fluctuation $\delta m_B$ in $MeV$ excited in various collisions. Main analysis A is the first set of numbers in each row, main analysis B is the second set of numbers.

| Energy | T | $\mu_B$ | $<m_B>$ | $\delta m_B$ | $C_{V,B}/B$ | $<m_B>$ | $\delta m_B$ | $C_{V,B}/(N_B+N_{\bar{B}})$ |
|---|---|---|---|---|---|---|---|---|
| 11.6 | 118.1 | 555 | 1061 | 277 | 7.0 | 1071 | 277 | 7.0 |
|  | 119.1 | 578 | 1094 | 286 | 7.3 | 1086 | 286 | 7.3 |
| 30 | 139.5 | 428.6 | 1295 | 593 | 19.6 | 1277 | 594 | 20.7 |
|  | 140.3 | 428. | 1316 | 612 | 20.6 | 1313 | 614 | 21.9 |
| 40 | 147.6 | 380.3 | 1513 | 853 | 34.9 | 1517 | 860 | 39.9 |
|  | 145.5 | 375.4 | 1426 | 770 | 29.5 | 1444 | 775 | 32.8 |
| 80 | 153.7 | 297.7 | 1771 | 1216 | 64.0 | 1856 | 1243 | 82.1 |
|  | 151.7 | 288.9 | 1629 | 1070 | 51.2 | 1715 | 1087 | 62.7 |
| 158 | 157.8 | 247.3 | 2113 | 1676 | 114.3 | 2331 | 1753 | 168.4 |
|  | 154.8 | 244.5 | 1804 | 1313 | 73.4 | 1952 | 1348 | 97.0 |

The contribution of decoupled mesons to $C_V$ can be obtained from eq(12). In ref[3]: $\pi's=1356, K's=160, K^{*'}s=66, \rho's=149, \eta=49, \omega=40$, are the multiplicity yields for the $80\,GeV*A\;\;Pb+Pb$ collision for the low lying well known mesons. Using eq(12), each of these mesons contributes to $C_V$ as follows: $C_{V,\pi}/N_\pi=13.2$, $C_{V,K}/N_K=28.8$, $C_{V,\eta}/N_\eta=32$, $C_{V,\omega}/N_\omega=49.5$, $C_{V,\rho}/N_\rho=48.5$, $C_{V,K^*}/N_{K^*}=59$. Thus mesons make a very large contribution to $C_V$ also.

## 3. Summary and Conclusions

The properties of the specific heat of hadronic matter produced in very high energy nucleus-nucleus collisions such as those at RHIC and CERN are studied in this paper. The grand canonical statistical model is used to develop expressions for the particle multiplicity distribution and energy caloric equation of state which is then used to obtain the specific heat. The constraints associated with baryon number and electric charge conservation are included to obtain an expression for the specific heat which contains the particle yields, the mass spectrum of produced particles, baryon number $B$ and overall charge $Z$ values, and three coefficients $C_{bb}$, $C_{qq}$ and $C_{bq}$. These coefficients come from the correlations introduced by requiring overall baryon number and electric charge conservation in the spectrum of produced particles. The coefficients also depend on the fundamental unit of charge being different for the case of 1/3 (Q-g phase) compared to 1 (hadron phase). $C_{bb}$ and $C_{qq}$ are also measures of baryon number and charge fluctuations [11,12]. The specific heat is not simply a sum of independent contributions arising from each type of particle. Rather $C_V$ has additional terms which significantly alter its value from this independent particle result. Resonance excitations allow for the possibility that individual particle yields change with $T$ and redistribute the conserved charge and baryon number on other particles. The behavior of $C_V$ is studied in some limiting cases to see how various quantities such as baryonic charge conservation, the production of anti-baryons and electric charge conservation affect it. The mass fluctuation in excited resonances is shown to play an important role. Using properties of the constraint equations, the specific heat and mass spectrum of excited hadrons is related to properties of the baryon and electric charge chemical potentials $\mu_B$ and $\mu_Q$. In particular, the specific heat is related to the curvature of the chemical potentials, its slope and its value. The mean hadronic mass that is produced in a heavy ion collision involves the chemical potentials and their slopes, and the mass fluctuation involves these quantities and the curvature of the chemical potentials. A recent parameterization of the baryonic chemical potential with $T$ is shown to lead to a very rapid increase in the baryonic component of the specific heat. If this expression correctly describes the behavior of $\mu_B$ near a limiting temperature $T_0$, then $C_V$ would diverge as $1/(T_0 - T)^2$. Moreover, the exponent 2 is independent of the functional form used near $T_0$. The presence of anti-baryons plays a key role in the temperature behavior of $C_V$, and this independence property.

Acknowledgments
The author would like to thank the DOE for its support of this research under grant number DOE: FG02-96ER-40987.